\documentclass[letter,nofootinbib,12pt]{revtex4-2}

\usepackage[colorlinks,citecolor=blue]{hyperref}
\usepackage{amsmath}
\usepackage{mathrsfs}
\usepackage{bbm}
\usepackage{amsfonts}
\usepackage{amssymb}
\usepackage{latexsym}
\usepackage{graphicx}
\usepackage[english]{babel}
\usepackage{multirow}
\usepackage{float}
\usepackage{url}
\usepackage{slashed}
\usepackage[compat=1.1.0]{tikz-feynman}
\usepackage{orcidlink}

\renewcommand{\Delta}{\varDelta} 
\renewcommand{\Gamma}{\varGamma} 
\renewcommand{\Omega}{\varOmega} 
\renewcommand{\Phi}{\varPhi} 
\renewcommand{\Psi}{\varPsi} 
\renewcommand{\Sigma}{\varSigma} 
\renewcommand{\Theta}{\varTheta} 
\renewcommand{\epsilon}{\varepsilon}

\newcommand{\be}{\begin{equation}}
\newcommand{\ee}{\end{equation}}
\newcommand{\ba}{\begin{array}}
\newcommand{\ea}{\end{array}}
\newcommand{\bea}{\begin{eqnarray}}
\newcommand{\eea}{\end{eqnarray}}

\usepackage{geometry}                		
\usepackage{graphicx}				
\usepackage{amssymb}
 
\begin{document}


\hfill March 2026

\bigskip
\bigskip

\begin{center}

{\bf \Large Additional TeV-Scale Particles \\
Predicted by Quartification }\\

\bigskip
\bigskip
\bigskip

{Paul H. Frampton}
\footnote{paul.h.frampton@gmail.com} 
\orcidlink{0000-0002-2219-0330} \\
{Dipartimento di Matematica e Fisica 'Ennio De Giorgi',\\
Universit`a del Salento and INFN-Lecce,\\
Via Arnesano, 73100 Lecce, Italy.}\\

\bigskip

{Thomas W. Kephart}
\footnote{tom.kephart@gmail.com} 
\orcidlink{0000-0001-6414-9590} \\
{Department of Physics and Astronomy,\\
 Vanderbilt University, \\
 Nashville, TN 37235, USA} \\

\bigskip

{\bf Abstract}

\end{center}

\bigskip

\noindent
The LHC has failed to discover any new elementary particle
since the Higgs boson completed the standard model in 2012,  
Here we adopt the attractive method of quiver gauge field
theories to make predictions of additional particles which might be found
at Run 4 of the upgraded LH-LHC scheduled to begin in 2030. We
use an $SU(3)^4$ quiver gauge theory and exhaustively classify all possibilities
according to how many of the added  states shlep, meaning acquire a 
super-heavy Dirac mass. We arrive at four different choices, each of
which suggests interesting positive outcomes for Run 4.

\newpage

\section{Introduction}

\begin{figure}[htpb]
    \centering
    \includegraphics[width=\textwidth]{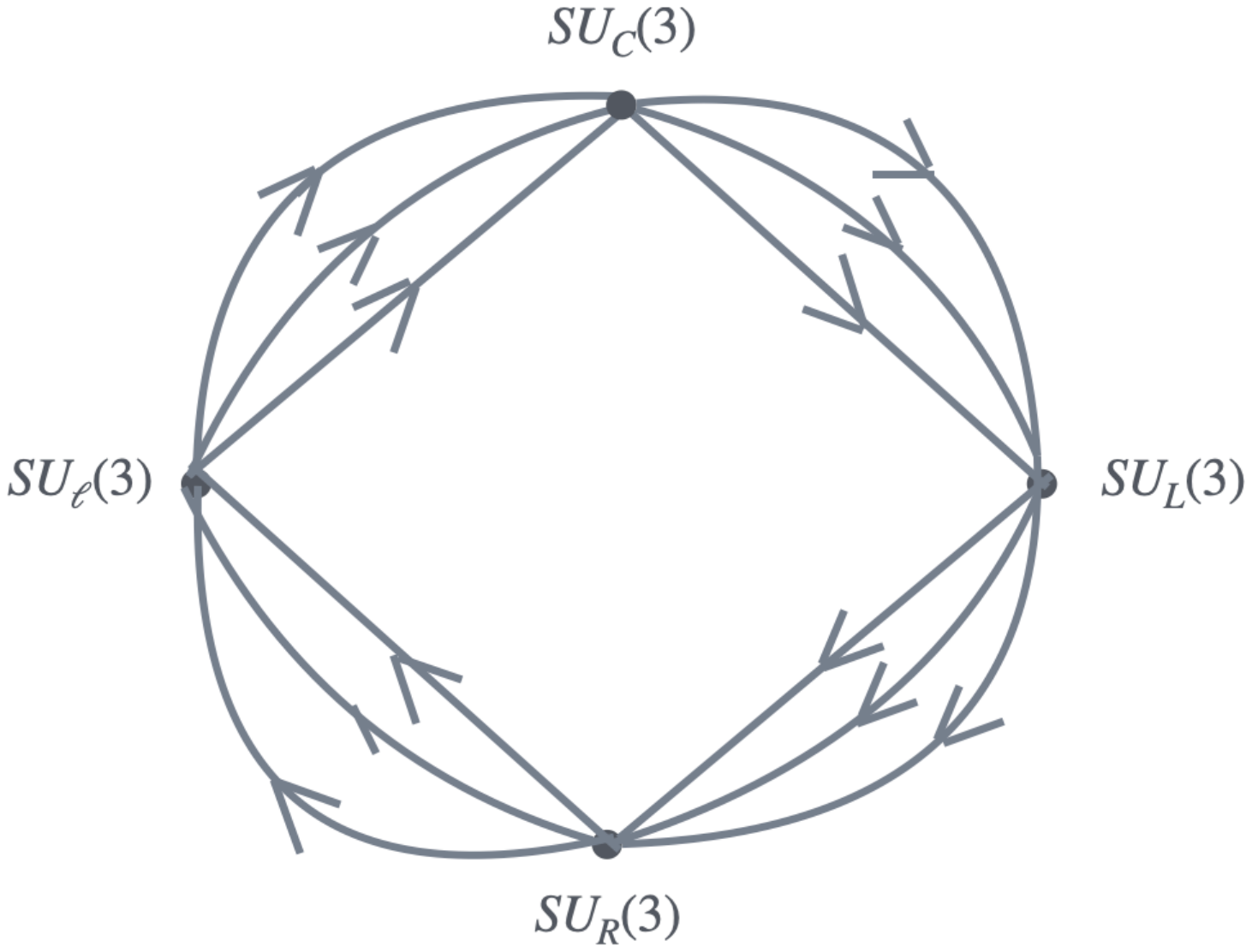}
    \caption{Quiver diagram for the quartification group with three families.}
    \label{quiver}
\end{figure}

\noindent
In order to make intelligent speculation about what will be the first
BSM particle to be discovered at the HL-LHC, we employ the
promising technique of quiver gauge theory, in particular we examine
the special case of quartification which has the gauge group $SU(3)^4$ \cite{JV,Babu:2003nw,Chen:2004jz,Babu:2007fx,Demaria:2005gk,Demaria:2006uu,Demaria:2006bd,Eby:2011ph}.
We assume a natural embedding of the SM and note that the starting
theory is anomaly-free by construction (see figure (\ref{quiver})), as are all theories arrived at
by its spontaneous symmetry breaking. We adopt the notation
of \cite{KephartWeiler}.   For a review of quiver gauge theories see \cite{Frampton:2007fr}.

\bigskip

\noindent
Our motivation is to analyse comprehensively the BSM particles
predicted at TeV scales.
It is sufficient first to analyse one family under the $SU(3)^4$ gauge
group (CLR$\ell$) which is
\begin{equation}
(3,3^*,1,1) + (1,3,3^*,1) + (1,1,3,3^*) + (3^*,1,1,3)
\label{family}
\end{equation}

\noindent
Initially at high energy the theory has quark-lepton symmetry  \cite{Foot:dw,Foot:fk,Foot:2006ie,FootVolkas}, but after spontaneous symmetry breaking only the color $SU(3)$ uniquely remains  exact. The other three SU(3)'s suffer SSB by
rank-preserving adjoint scalar VEVs as follows
\begin{equation}
SU(3)_L \rightarrow SU(2)_L \times U(1)_A 
\end{equation}
\begin{equation}
SU(3)_R \rightarrow U(1)_B \times U(1)_C
\end{equation}
\begin{equation}
SU(3)_{\ell} \rightarrow SU(2)_{\ell} \times U(1)_D
\end{equation}

\bigskip

\noindent
We define the normalisations of the Gell-Mann matrices which represent
the Cartan subalgebras of the three non-color SU(3)'s as
\begin{equation}
A, C, D \sim (1,1,-2)
\end{equation}
\begin{equation}
B \sim (1,-1,0)
\end{equation}
and note that A,B,C,D change sign under the complex conjugation $3 \rightarrow 3^*$.

\bigskip

\noindent
The weak hypercharge $Y$ in the charge embedding $Q=T_3 + Y$ is defined by
\begin{equation}
Y = -\frac{1}{6}A + \frac{1}{2}B -\frac{1}{6}C + \frac{1}{2}D
\label{WeakHypercharge}
\end{equation}

\bigskip

\noindent
With these conventions we can express the four terms in Eq.(\ref{family}) first in the form $(CL{\ell})_{A,B,C,D}$
then, using Eq.(\ref{WeakHypercharge}), in the more convenient form $(CL{\ell})_Y$  as follows

\begin{eqnarray}
(3,3^*,1,1) & \rightarrow & (3,2,1)_{-1,0,0,0} + (3,1,1)_{2,0,0,0} \nonumber \\
& \rightarrow & (3,2,1)_{\frac{1}{6}} +  (3,1,1)_{-\frac{1}{3}} \nonumber \\
\label{first}
\end{eqnarray}

\begin{eqnarray}
(1,3,3^*,1) & \rightarrow & (1,2,1)_{1,-1,-1,0} + (1,2,1)_{1,1,-1,0} + (1,2,1)_{1,0,2,0} + \nonumber \\
& & +  (1,1,1)_{-2,-1,-1,0} + (1,1,1)_{-2,1,-1,0} + (1,1,1)_{-2,0,2,0} \nonumber \\
& \rightarrow & (1,2,1)_{-\frac{1}{2}} +  (1,2,1)_{+\frac{1}{2}} +(1,2,1)_{-\frac{1}{2}} + \nonumber \\
& & +  (1,1,1)_0 + (1,1,1)_{+1} + (1,1,1)_{0} \nonumber \\
\label{second}
\end{eqnarray}

\begin{eqnarray}
(1,1,3,3^*) & \rightarrow & (1,1,2)_{0,1,1,-1} + (1,1,2)_{0,-1,1,-1} + (1,1,2)_{0,0,-2,-1} + \nonumber \\
& & + (1,1,1)_{0.1.1.2} + (1,1,1)_{0,-1,1,2} + (1,1,1)_{0,0,-2,+2} \nonumber \\
& \rightarrow & (1,1,2)_{0} +  (1,1,2)_{-1}  + (1,1,2)_0 +  \nonumber \\
& & + (1,1,1)_1 + (1,1,1)_0 + (1,1,1)_1. \nonumber \\
\label{third}
\end{eqnarray}

\begin{eqnarray}
(3^*,1,1,3)) & \rightarrow & (3^*,1,2)_{0,0,0,1} + (3^*,1,1)_{0,0,0,-1} \nonumber \\
& \rightarrow & (3^*,1,2)_{\frac{1}{3}} +  (3^*,1,1)_{-\frac{1}{3}} \nonumber \\
\label{fourth}
\end{eqnarray}

\bigskip

\noindent
Reverting to the conventional standard model notation $(3_C,2_L)_Y$,
we can confirm that among the 36 states of 
Eq.(\ref{family}) and appearing in Eqs (\ref{first},\ref{second},\ref{third},\ref{fourth}) there is 
one normal family of 15 states

\begin{equation}
(3,2)_{\frac{1}{6}} +(3^*,1)_{\frac{1}{3} } + (3^*,1)_{-\frac{2}{3}} + (1,2)_{-\frac{1}{2}} + (1,1)_{+1}
\label{SMfamily}
\end{equation}

\bigskip

\noindent
We divide the remaining 21 states into the three categories\\
(1) Six shleppable quarks
\begin{equation}
(3,1)_{-\frac{1}{3}} + (3^*,1)_{+\frac{1}{3}}
\label{ShleppableQuarks}
\end{equation}
(2) Eight shleppable charged leptons
\begin{equation}
(1,2)_{\frac{1}{2}} + (1,2)_{-\frac{1}{2}} +2(1,1)_1 + 2(1,1)_{-1}
\label{ShleppableChargedLeptons}
\end{equation}
(3) Seven sterile neutrinos
\begin{equation}
7 (1,1)_0
\label{SterileNeutrinos}
\end{equation}

\bigskip

\noindent
After these prefatory remarks, we shall analyse different possibilities
according to the degree of shlepping {\it i.e.,} the number of states in these
categories which are shlepped to a mass scale well over a TeV, that is
to at least 10 TeV or much higher. We shall assume that the TeV physics
depends primarily only on  the BSM states which are not shlepped.

\noindent
\section{Maximal Shlepping (Quarks and Charged Leptons)}

\noindent
In this case the BSM states predicted at the TeV scale are, in $(3_C,2_L)_Y$
notation, just
\begin{equation}
7(1,1)_0.
\label{MaximalShlepping}
\end{equation}

\bigskip

\noindent
This Eq.(\ref{MaximalShlepping}) displays an exceptionally inert
set of seven Weyl states without strong or electroweak, only
gravitational, interactions which are thereby impractical to detect
in any terrestrial experiment.

\bigskip

\noindent
It is very interesting that such sterile neutrinos are confidently
predicted. They are similar to the axion-like-paricles (ALPs)
predicted by string theory which is based on richer assumptions.
They could be related to dark matter whose only established
interaction is gravitational.

\bigskip

\noindent
These Weyl spinors, 21 including all three families, could also
play a r\^{o}le in the see-saw mechanism\cite{Minkowski} for neutrino masses.

\bigskip

\noindent
The states in Eq.(\ref{MaximalShlepping}) can acquire Majorana
masses by including explicitly in the lagrangian the terms
\begin{equation}
\Sigma M_{ij} \Psi^i_L \Psi^J_L
\label{Majorana}
\end{equation}
where $M_{ij}$ is symmetric. Global fermion number symmetries of an unfamiliar type
are violated by these mass terms.

\noindent
\section{Partial Shlepping (I) (Charged Leptons Only)}

\noindent
In this case there are extra quark states at TeV scales given
in Eq, (\ref{ShleppableQuarks})
\begin{equation}
(3,1)_{-\frac{1}{3}} + (3^*,1)_{+\frac{1}{3}}
\label{ShleppableQuarks2}
\end{equation}
for each of the three families
\begin{equation}
(\hat{d})_{L+R} + (\hat{s})_{L+R}+(\hat{b})_{L+R}
\label{DownLike}
\end{equation}
which describe three new  down-type quark each
with a Dirac mass term.

\bigskip

\noindent
The extra
$Q=-\frac{1}{3}$ quark in each family will, in general,
mix with the known quark of the same charge. The new quark
can acquire a Dirac mass because all the four components of a
Dirac spinor are present in Eq.(\ref{ShleppableQuarks}).

\bigskip

\noindent
Quantum mechanics dictates that a new $\hat{d}$ quark can
mix with the already known $d$ quark and hence violate
unitarity of the usual $3 \times 3$ CKM matrix. We can
keep as the flavor eigenstates the mass eigenstates
of (u, c, t) quarks. The generalisation of CKM in the SM
will now involve the very interesting generalisation to
flavor eigenstates which are mixtures of
$(d, \hat{d}; s, \hat{s}; b, \hat{b})$.

\bigskip

\noindent
The additional TeV scale quarks will contribute virtually
to weak decays into light quarks, so that precision
measurements of all such decays need to be made
to place upper bounds on the new mixing angles between
$d-\hat{d}$, $s - \hat{s}$, and $b - \hat{b}$.

\bigskip

\noindent
Quite apart from the possible smallness of mixing angles,
there is another suppression caused by the virtual
unshlepped states in Feynman diagrams of the
type $\sim M(d) / TeV$ and a rich phenomenology
ensues.

\bigskip

\noindent
We  shall defer detailed calculations of all the
fascinating physics to future publications. The method
of choice would be to use the powerful techniques of effective
field theory such as by employing the Standard Model Effective Field Theory (SMEFT) framework, see e.g. \cite{Isidori:2023pyp}.

\noindent
\section{Partial Shlepping (II) (Quarks Only)}

\noindent
In this case there are extra charged lepton states at TeV scales given
in Eq, (\ref{ShleppableChargedLeptons})
\begin{equation}
(1,2)_{\frac{1}{2}} + (1,2)_{-\frac{1}{2}} +2(1,1)_1 + 2(1,1)_{-1}
\label{ShleppableChargedLeptons2}
\end{equation}
which describe an $SU(2)_L$ doublet and singlet, all with Dirac
mass terms.
\begin{equation}
(   N , E^- )_{L+R} + (E^+) _{L+R}
\label{SU2chargedleptons}
\end{equation}

\bigskip

\noindent
If the  charged leptons remain unshlepped, there is a fascinating
situation where the Dirac spinors present in 
Eq.(\ref{ShleppableChargedLeptons}) add extra 
\begin{equation}
\hat{e^{\pm}}, \hat{\mu}^{\pm}, \hat{\tau}^{\pm}
\label{Addedchargedleptons}
\end{equation}
states, as well as extra lepton doublets
\begin{equation}
(\hat{\hat{\nu_e}}, 
\hat{\hat{e^-}}) + 
(\hat{\hat{\nu_{\mu}}}, 
\hat{\hat{\mu^-}}) + 
(\hat{\hat{\nu_{\tau}}},
 \hat{\hat{\tau^-}})
\label{AddedLeptonDoublets}
\end{equation}
which can mix with their SM counterparts. This will
lead to non-unitarity of the PMNS matrix. 

\bigskip

\noindent
A phenomenological analysis using SMEFT is urgently needed to put limits on this
potential violation of unitarity.
 
 \bigskip

\noindent
\section{Minimal Shlepping (All Additional States at TeV Scales)}

\noindent
Minimal shlepping will lead to an ``embarrassment of riches" scenario with all the states
of the previous two sections at TeV scales. This implies a full
twenty-one additional states per family and non-unitarity of {\it both}
CKM {\it and} PMNS mixing matrices. We regard this as the least
probable scenario and have ordered our scenarios from the
most likely to the least likely, although this is purely aesthetic
and all are perfectly possible.

 \newpage

\noindent
\section{Discussion}
\noindent
Quiver gauge theories are one of the most attractive ways to
propose BSM model building and, most importantly, to make
plausible predictions about BSM particles. This is not least
because the SM itself is essentialy a quiver gauge theory.
Cancellation of all triangle anomalies arising from the chiral
bifundamental fermions is guaranteed by the quiver construction.

\bigskip

\noindent
Of quiver theories, in this article we have focused on the
specific example of $SU(3)^4$ which has the small order $g=32$
and where each family contains 36 Weyl states, compared to
the 15 of the SM, and therefore 21 possible predictions for
new physics BSM particles. We have analysed as comprehensively
as possible which are most plausible as TeV scale particles.

\bigskip

\noindent
We have employed the useful notion of shlepping\footnote{To our knowledge,
this Yiddish addition to the particle theory lexicon to describe superheavy
Dirac masses was originally due to Glashow at some time in the 1970s.} systematically 
to analyse four
different scenarios for the TeV BSM particles predicted by our version
of quartification theory.

\bigskip

\noindent
The 21 additional chiral states occurring in each family naturally divide
into quarks (6), charged leptons (8) and sterile neutrinos (7).
If all the additional quarks and charged leptons are shlepped,
there remain as BSM particles only the fascinating seven
inert states at TeV scales, or conceivably much higher masses.

\bigskip

\noindent
These inert states interact only gravitationally and therefore
have at least two obvious applications :
\begin{itemize}
\item As candidates for microscopic constituents of dark matter.
\item As superheavy right-handed neutrinos participating in the
see-saw  mechanism for generating neutrino masses.
\end{itemize}

\bigskip

\noindent
As dark matter candidates, they can play the r\^{o}le ascribed  to
ALPs in string cosmology without any of the rich mathematical
superstructure concomitant with the description of fundamentsl
strings.

\bigskip

\noindent
In their second possible r\^{o}le, these inert states could lie at
a natural value $\sim 10^{10}$ GeV for the see-saw mass
usually assumed in neutrino theory.

\bigskip

\noindent
If the additional quarks remain unshlepped there is an extra
$Q=-\frac{1}{3}$ quark in each family which  will, in general,
mix with the known quark of the same charge. The new quark
can acquire a Dirac mass because all the four components of a
Dirac spinor are present in Eq.(\ref{ShleppableQuarks}).
A further complication is that the new $\hat{d}$ quark can
mix with the already known $d$ quark and hence compromise
the unitarity of the $3 \times 3$ CKM matrix. It would be
interesting to put phenomenological limits on this new
type of mixing.

\bigskip

\noindent
If the  charged leptons remain unshlepped, there is a similar
situation where the Dirac spinors present in 
Eq.(\ref{ShleppableChargedLeptons}) add extra $\hat{e^{\pm}}$
states, as well as extra $(\hat{\nu_e}, \hat{e^-})$
states which can mix with their SM counterparts. This would
lead to non-unitarity of the $3 \times 3$ PMNS matrix. Again,
phenomenological analysis is needed to put limits on this
potential violation of unitarity.

\bigskip

\noindent
In our final scenario, we have discussed when all the
additional quark and charged lepton states are at 
TeV scales.  This will lead to unitarity issues for {\it both}
the CKM {\it and} the PMNS mixing matrices, and a
very rich phenomenology.

\bigskip

\noindent
To conclude, we hope to have been convincing that
our quartification model provides predictions of
many potential BSM particle candidates at TeV scales, and an
almost overwhelming
number of interesting phenomenological questions.

\end{document}